\documentclass[pra
	       ,nofootinbib
	       ,floatfix
	       ,superscriptaddress
	       ,twocolumn
	       ]{revtex4-1}

\usepackage{times}
\usepackage{mathtools}
\usepackage{blkarray}

\usepackage{soul}
\usepackage{amsmath} 
\usepackage{mathrsfs}
\usepackage{amssymb}
\usepackage{cool}
\usepackage{graphicx} 
\usepackage{color} 
\usepackage[utf8]{inputenc}
\usepackage[hidelinks]{hyperref}
\usepackage[hidelinks]{hyperref}
\usepackage{multirow}
\usepackage{siunitx}
\usepackage[english]{babel}

\newcommand{\SOMfirstPICposition}{\begin{figure}[b]}

\begin{document}

\title{Analytical Pendulum Model for a Bosonic Josephson Junction}

\author{Marine~Pigneur}
\affiliation{Vienna Center for Quantum Science and Technology, Atominstitut, TU Wien, Stadionallee 2, 1020 Vienna, Austria}
 
\author{J\"org~Schmiedmayer}
\affiliation{Vienna Center for Quantum Science and Technology, Atominstitut, TU Wien, Stadionallee 2, 1020 Vienna, Austria}

\date{\today}

\begin{abstract}
We present an analytical description of the tunneling dynamics between two coupled Bose-Einstein condensates in the Josephson regime. The model relies on the classical analogy with a rigid pendulum and focuses on two dynamical modes of this system: Josephson oscillations and Macroscopic Quantum Self-Trapping. The analogy is extended to include an energy difference between the two superfluids caused by an asymmetry in the trapping potential. The model is compatible with the mean-field predictions of the two-mode Bose-Hubbard model. It gives new insights on the mean-field model by involving experimentally measurable parameters with reduced correlations. For agreement with recent experimental observations, we establish heuristic formulas including a dissipation. We conclude with a convincing application of the model to several sets of experimental results.

\end{abstract}

\maketitle



\section{Introduction}

The physics of quantum many-body systems out-of-equilibrium is a very challenging problem of modern physics. In recent years, the field has experienced rapid progress partly attributed to the unprecedented insights provided by ultra-cold atoms. They constitute versatile and highly controllable systems ideal to address such questions \cite{langen_ultracold_2015} and their theory have been studied extensively \cite{rigol_relaxation_2007,polkovnikov_colloquium_2011,eisert_quantum_2015}.
 
This is well illustrated by a Bosonic Josephson Junction (BJJ) consisting of two coupled superfluids. The dynamics of a BJJ has been observed experimentally by \cite{albiez_direct_2005,levy_2007,leblanc_dynamics_2011,spagnolli_crossing_2017,pigneur_relaxation_2018}. However, many questions remain to this day, among which the unexplained damping of the dynamics observed by \cite{leblanc_dynamics_2011,pigneur_relaxation_2018} for a BJJ close or in the one-dimensional (1D) regime. In \cite{pigneur_relaxation_2018}, the two elongated 1D quasi-condensates are initialized with a phase difference. After few oscillations, the phase difference and atom number difference relax to a phase-locked equilibrium state without identified mechanism. The feature is observed on every single experimental realization and does not result from an ensemble averaging. 

The sine-Gordon model \cite{gritsev_linear_2007} has proven to be very successful in describing the equilibrium dynamics of two coupled 1D atomic superfluids up to very high order correlations \cite{schweigler_experimental_2017}. However, its suitability to predict the dynamics out-of-equilibrium is questioned by \cite{pigneur_relaxation_2018}. Indeed, theoretical approaches such as quenches in the sine-Gordon model using exact solutions in the single-mode approximation, truncated Wigner approximation and variational Gaussian approaches fail in reproducing the fast phase-locking \cite{dalla_torre_universal_2013,priv_comm}. 

The elongated dimension introduces a spectrum of excitation modes whose population is responsible for phase fluctuations. The phase fluctuations reported in \cite{pigneur_relaxation_2018} are very small, such that only the dynamics of the lowest energy mode could be studied. While it does not exclude a coupling between longitudinal modes not appearing on the measured observables, it reduces the complexity of the data analysis to a zero dimensional (0D) treatment. The two-mode Bose-Hubbard model (TMBH) describes accurately BECs in a 0D double-well \cite{gati_bosonic_2007}. It has been investigated extensively using various approaches, such as the  Gross-Pitaevskii approximation \cite{leggett_bose-einstein_2001}, the mean-field theory \cite{milburn_quantum_1997}, the quantum phase model \cite{anglin_exact_2001} and the Bethe ansatz method \cite{zhou_exact_2003}. More recent models go further, presenting approaches beyond mean-field \cite{ananikian_gross-pitaevskii_2006,jia_nonlinear_2008,sakmann_exact_2009,sabin_analytical_2014}, beyond the two-mode approximation \cite{gillet_tunneling_2014} or beyond the linear tunneling \cite{rubeni_two-site_2017}. 

Models predict various phenomena leading to the damping of the dynamics of the BJJ. It was predicted very early that the mean-field dynamics would be modulated by quantum collapses and revivals \cite{milburn_quantum_1997}. More recently, a quantum collapse caused by an exponential growth of the quantum phase dispersion is expected for large atomic imbalance between the two wells \cite{shchesnovich_fock-space_2008}. In the case of an open system, the interaction with a thermal bath is proposed in \cite{zapata_josephson_1998,savenko_stochastic_2013,rajagopal_dynamics_2015}. A damping can also be attributable to phase noise and particle dissipation \cite{trimborn_mean-field_2008,ferrini_noise_2010,li_phase_2017}. An energy loss through excitations to higher energy transverse modes is discussed in \cite{gillet_tunneling_2014} and very recently in \cite{lappe_fluctuation_2018,polo_damping_2018}.

All these models affect measurable observables and can therefore be tested. The atomic losses are excluded in \cite{pigneur_relaxation_2018} as the atom number measured with a single-atom sensitivity \cite{bucker_single-particle-sensitive_2009} is constant. Other models predict damping and revival of the dynamics but at times much longer than the relaxation, where the physics of the system is dominated by its 1D character \cite{bouchoule_modulational_2005}. Other predictions rely on excitations to higher energy transverse states. The measurement of the relative phase by atomic interferometry would show this as a degradation of the fringes contrast, which is not observed.
  
As no microscopic origin of the damping could be identified in \cite{pigneur_relaxation_2018}, the analysis of the observations relies on a phenomenological damping term added to the mean-field TMBH as in \cite{marino_bose-condensate_1999,zheng_dissipation_2012}. While this approach describes the damping, it is not ideal to fit the data. The parameters involved show strong correlations which make the fit results unstable. Additionally, the parameters are not directly accessible experimentally, making them difficult to check.

For these reasons, we extend in \cite{pigneur_relaxation_2018} the classical analogy with a mechanical pendulum presented in \cite{marino_bose-condensate_1999} and establish the analytical oscillating solutions of the pendulum applied to the mean-field TMBH. This introduces parameters with small correlations and directly accessible experimentally \cite{pigneur_relaxation_2018}. It also provides new insights on the TMBH model.

In the present work, we present the analytical solutions of the BJJ in the entire Josephson regime by extending the solutions mentinoned in \cite{raghavan_coherent_1999,liang_dynamics_2010}. We go beyond the pendulum analogy to treat an asymmetric BJJ. In a second part, we empirically introduce a friction as in \cite{marino_bose-condensate_1999} and establish the heuristic solutions of a dissipative BJJ. In a last section, we present a successful application of the model to experimental results and illustrate the physical insights gained by the model.

\section{Dynamics of a Bosonic Josephson Junction}
For two Bose-Einstein condensates (BECs) trapped in a double-well potential, we define the atom number imbalance normalized by the total atom number:
\begin{equation}
n=\frac{N_L-N_R}{N_L+N_R},
\end{equation}
with $N_{L,R}$ the atom number of the left and right component, respectively. The conjugated variable is the relative phase, defined by:
\begin{equation}
\phi=\phi_L-\phi_R,
\end{equation} 
with $\phi_{L,R}$ the phase of the left and right component, respectively. Phase and imbalance are accessible experimentally \cite{pigneur_relaxation_2018} and the mean-field TMBH model predicts that they evolve over time for any initial state whose phase or imbalance differs from zero.
\begin{figure}[h] 
\includegraphics[]{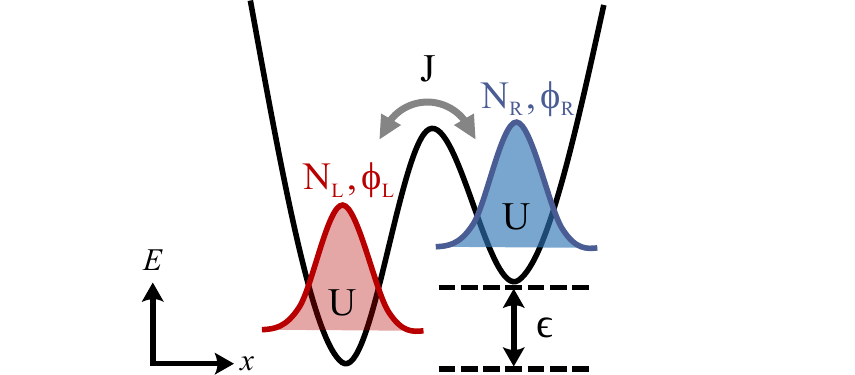}
\caption{\label{Fig:BJJ_schematics} Schematics of the Bosonic Josephson Junction consisting of two sites (left $L$ and right $R$). The BECs are characterized by an atomic population $N_{L,R}$ and a phase $\phi_{L,R}$. The BECs have an on-site interaction energy $U$ and are coupled with a single-particle tunnel energy $J$. A trap asymmetry translates in a detuning energy $\epsilon=E_L-E_R$. }
\end{figure}

\subsection{Dynamical modes of the symmetric BJJ in the mean-field two-modes Bose-Hubbard model}
The mean-field TMBH model involves three energy contributions: the on-site interaction energy $U$ that we consider identical on the left and right site, the single-particle tunnel-coupling energy  $J$ and a detuning $\epsilon=E_L-E_R$ between the trap minima [Fig.~\ref{Fig:BJJ_schematics}]. In this model, one can derive the equations of motion to establish the dynamical modes of the system \cite{smerzi_quantum_1997,raghavan_coherent_1999}:
\begin{align}
\dot{n}(t)&=-\frac{2J}{\hbar}\sqrt{1-n^2(t)}\sin\phi(t), \label{eq:n_dot_sym}\\
\dot{\phi}(t)&=\frac{\epsilon}{\hbar} + \frac{2J}{\hbar}\left[\Lambda n(t)+\frac{n(t)}{\sqrt{1-n^2(t)}}\cos\phi(t)\right]. \label{eq:phi_dot_sym}
\end{align}
$\Lambda$ characterizes the interplay between the inter-atomic interaction and the tunneling. It is defined by ${\Lambda=NU/2J}$ with $N=N_L+N_R$ the total atom number.

For a symmetric BJJ ($\epsilon=0$), the dynamical modes are fully determined by $\Lambda$ and by the initial conditions ($n_0,\phi_0$). They define a parameter $\alpha$, constant at all times and defined by:  
\begin{equation}
\alpha=\frac{\Lambda}{2}n_0^2-\sqrt{1-n_0^2}\cos(\phi_0) \text{ with } \alpha\in [-1,\frac{\Lambda}{2}].
\label{eq:separatrix}
\end{equation}

\begin{figure}[t] 
\includegraphics[]{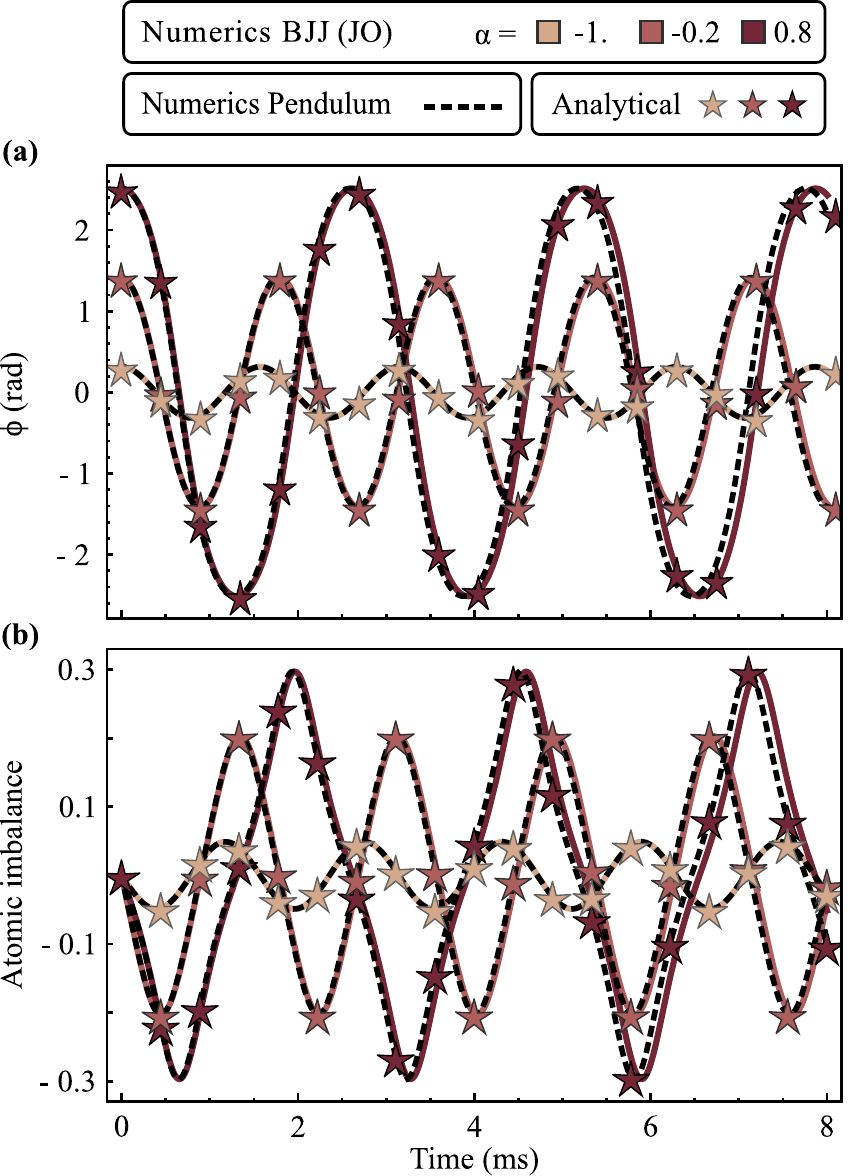}
\caption{\label{Fig:JO_BJJ} Oscillating dynamics of the phase (a) and atomic imbalance (b) of $N=\SI{5000}{\text{atoms}}$ for three initial states: $n_0=0$ and $\phi=[0.1,0.45,0.8]\pi$. The TMBH parameters are ${J=\SI{50}{\hbar\cdot \mathrm{Hz}}}$, ${U=\SI{0.8}{\hbar\cdot \mathrm{Hz}}}$ such that ${\alpha=-1,-0.2}$ and $0.8$. For ${\alpha \rightarrow 1}$, the oscillations present an increasing anharmonicity and a decreasing frequency. We represent the numerical solutions of the equations of motion of the mean-field TMBH model (color plain lines), the numerical solutions of the corresponding pendulum (black dashed line) and its analytical solutions (color stars). For ${\alpha \approx 1}$, the pendulum frequency differs slightly from the TMBH predictions.} 
\end{figure}

For ${-1<\alpha<1}$, the dynamics present Josephson oscillations \cite{javanainen_oscillatory_1986,jack_coherent_1996,zapata_josephson_1998}. The phase and imbalance oscillate $\pi$ out of phase. At small amplitudes (${-1<\alpha \ll 1}$), the oscillations are harmonic and oscillate with the plasma frequency $\omega_0$. Larger amplitude oscillations (${\alpha \lesssim 1}$) are anharmonic and their frequency decreases compared to $\omega_0$ (Fig.~\ref{Fig:JO_BJJ}). 

\begin{figure}[t] 
\includegraphics[]{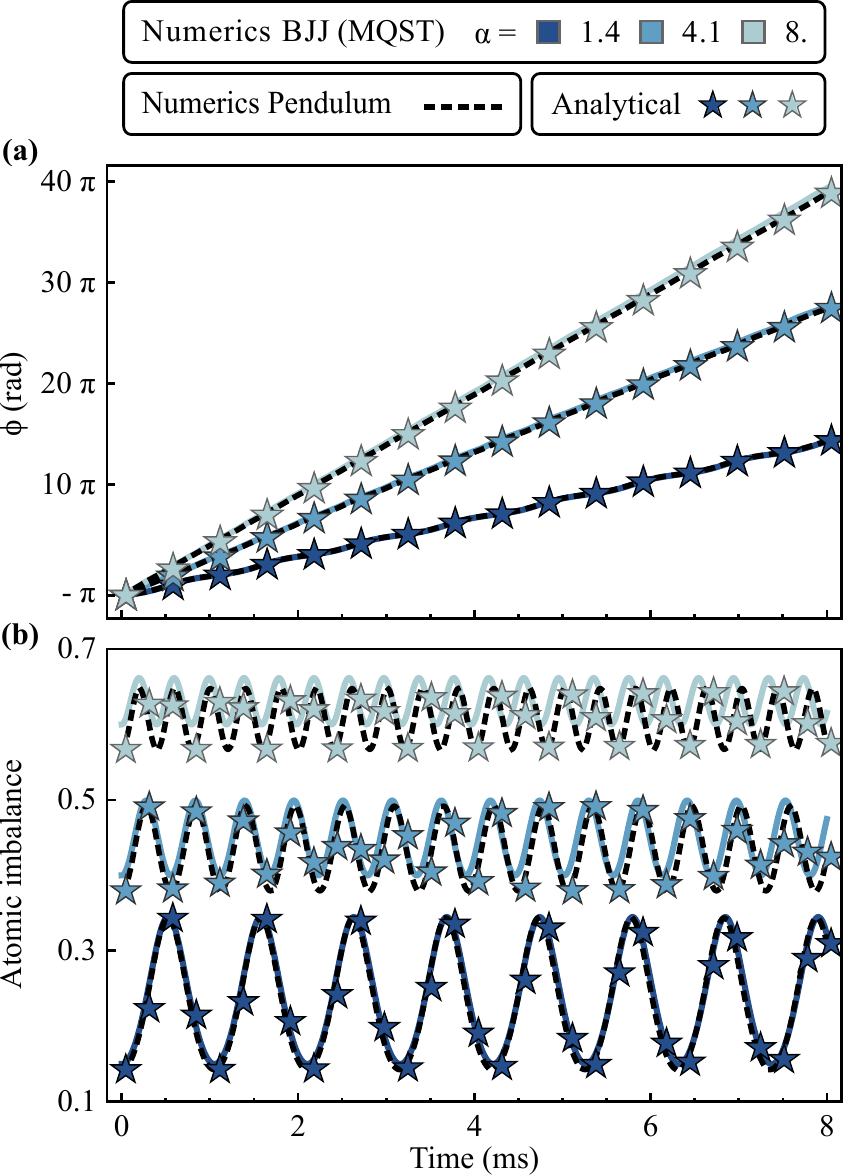}
\caption{\label{Fig:ST_BJJ} Self-trapped dynamics of the phase (a) and atomic imbalance (b) of $N=\SI{5000}{\text{atoms}}$ for three initial states: $\phi_0=-\pi$ and $n_0=[0.15,0.4,0.6]$. The TMBH parameters are ${J=\SI{50}{\hbar\cdot \mathrm{Hz}}}$, ${U=\SI{0.8}{\hbar\cdot \mathrm{Hz}}}$ such that ${\alpha=1.4,4.1}$ and $8$. The phase accumulation rate and mean imbalance increase with $\alpha$. The period and amplitude of the imbalance oscillation decrease with $\alpha$. The  color lines are the numerical solutions of the mean-field TMBH model, the black dashed lines are the numerical solutions of the corresponding pendulum the stars are its analytical solutions. For $\alpha \gg 1$, the pendulum results differ slightly from the TMBH predictions.} 
\end{figure}

For ${1<\alpha<\Lambda/2}$, the system is in the Macroscopic Quantum Self-Trapping (MQST) regime. The imbalance exhibits small oscillations around an averaged value $\text{sign}(n_0)\bar{n}$ with $0<\bar{n}< 1$. The averaged value of the imbalance and its oscillations frequency, as well as the phase accumulation rate, increase with $\alpha$ (Fig.~\ref{Fig:ST_BJJ}). 

\begin{figure}[t] 
\includegraphics[]{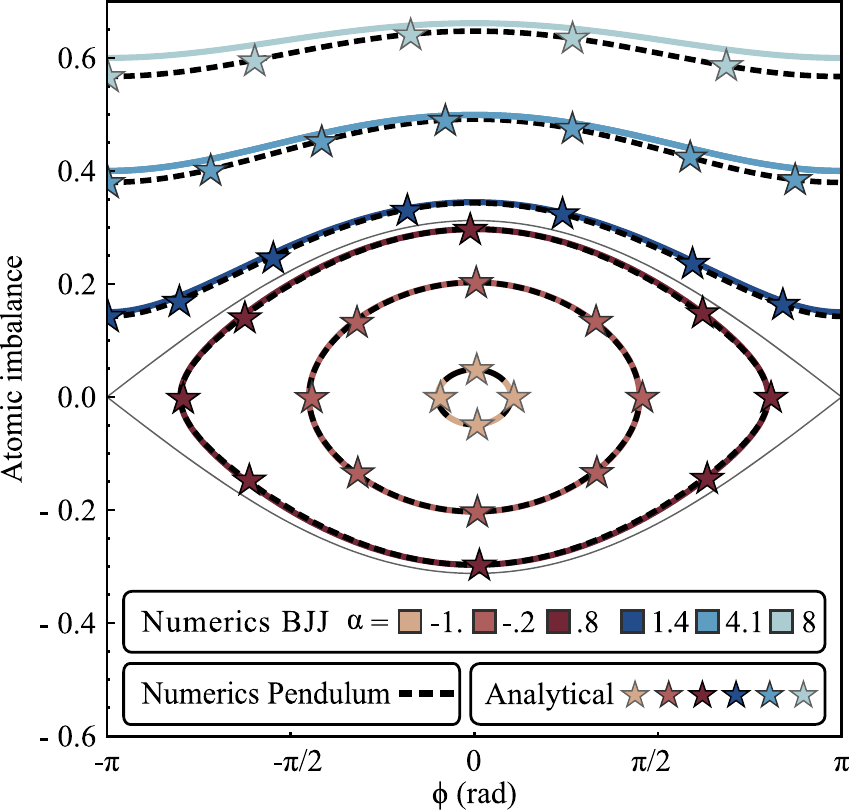}
\caption{\label{Fig:Phase_portrait} Phase portrait of the dynamics of $N=\SI{5000}{\text{atoms}}$ for the six initial states introduced in Fig.~\ref{Fig:JO_BJJ} and Fig.~\ref{Fig:ST_BJJ} and for the separatrix (grey line). The system parameters are ${J=\SI{50}{\hbar\cdot \mathrm{Hz}}}$ and ${U=\SI{0.8}{\hbar\cdot \mathrm{Hz}}}$. The numerical solutions of the mean-field TMBH are the color plain lines, the numerical solutions of the pendulum for comparable parameters are the black dashed line and its analytical solutions are the color stars. It shows clearly that the TMBH and the pendulum gives identical amplitudes in the oscillating regime and differs only slightly deep in the MQST.} 
\end{figure}

$\alpha=-1$ is the stable equilibrium point such the phase and imbalance remains zero at all time. $\alpha=1$ is the threshold value between the oscillating and self-trapped regimes. It defines the separatrix such the system evolves toward the unstable equilibrium point $n=0$ and ${\phi=\pm \pi}$ (Fig.~\ref{Fig:Phase_portrait}).  

\subsection{Analogy with a rigid pendulum}

The symmetric BJJ is analoguous to a classical pendulum of imbalance-dependent length ${l(t)=\sqrt{1-n^2(t)}}$ \cite{raghavan_coherent_1999,marino_bose-condensate_1999,leggett_bose-einstein_2001}. The Josephson oscillations of the TMBH model are analogous to the oscillating motion of a classical pendulum (Fig.~\ref{Fig:JO_BJJ}), while the MQST is analogous to the full-swing of the pendulum (Fig.~\ref{Fig:ST_BJJ}). We restrict this study to the rigid pendulum, which translates to the approximation ${n(t)\ll 1}$. 

The pendulum-like differential equation for the phase evolution is obtained by combining Eq.~(\ref{eq:n_dot_sym}) and the time derivative of Eq.~(\ref{eq:phi_dot_sym}): 
\begin{equation}
\ddot{\phi}(t)+\omega_0^2\sin(\phi(t))=0.
\label{eq:phi_dot_dot}
\end{equation}
$\omega_0$ is the plasma frequency defined by:
\begin{equation}
\omega_0=\frac{2J}{\hbar}\sqrt{\Lambda+\lambda},
\label{eq:omega_0}
\end{equation}
with $\lambda=\cos(\phi_0)$ often approximated to $1$. To solve Eq.~(\ref{eq:phi_dot_dot}) numerically, one must know $\Lambda,J$ and the initial conditions $\phi_0$ and $\dot{\phi}_0$ given by Eq.~(\ref{eq:phi_dot_sym}).

The time evolution of the imbalance is deduced from Eq.~(\ref{eq:n_dot_sym}) by:
\begin{equation}
n(t)\approx\frac{\hbar}{2J}\frac{1}{\Lambda + \lambda}\dot{\phi}(t).
\label{eq:n_pend}
\end{equation}
 
\subsection{Analytical solutions of the pendulum applied to the BJJ}\label{sec:analytical_sol_no_eta}

In analogy with the pendulum, the energy function $E$ of the symmetric BJJ is a conserved quantity given by 
\begin{equation}
E=\dot{\phi}^2(t)+4\omega_0^2 \sin^2\left(\frac{\phi(t)}{2}\right),
\label{eq:E}
\end{equation}
with $\phi(t)$ the phase and $\omega_0$ the plasma frequency \cite{marion_classical_2004}. In particular $E$ at the initial time reads  
\begin{equation}
E_0=\dot{\phi}_0^2+4\omega_0^2\sin^2\left( \frac{\phi_0}{2}\right),
\label{eq:E0}
\end{equation}
with $\dot{\phi}_0$ given by Eq.~(\ref{eq:phi_dot_sym}):
\begin{equation}
\dot{\phi}_0=\frac{2J}{\hbar} \left[\Lambda + \lambda\right] n_0.
\label{eq:phi_dot0}
\end{equation}

We express $\dot{\phi}(t)$ using Eqs.~(\ref{eq:E},\ref{eq:E0}) to obtain the differential equation for the phase:
\begin{equation}
\dot{\phi}(t)^2=E_0-4\omega_0^2\sin^2\left( \frac{\phi(t)}{2}\right).
\label{eq:phidot^2}
\end{equation}

We perform the change of variable $y(t)=\phi(t)/2$ and ${\tilde{t}=k\omega_0 t}$ where $k^2$ is the ratio of the total energy to the maximal potential energy:
\begin{equation}
k^2=\frac{E_0}{4\omega_0^2}.
\label{eq:k}
\end{equation} 
Eq.~(\ref{eq:phidot^2}) becomes:
\begin{equation}
\frac{dy}{d\tilde{t}}=\sqrt{1-k^{-2}\sin^2(y)},
\label{eq:y_dot}
\end{equation}
and is solved for $\phi_0=0$ by the Jacobi amplitude function $\JacobiAmplitude{y}{k^{-2}}$ \cite{abramowitz_handbook_1972}. This definition of the Jacobi amplitude adopts the convention of \cite{abramowitz_handbook_1972} and of Wolfram Mathematica\texttrademark.
We deduce that the evolution of the relative phase in a BJJ reads:
\begin{equation}
\phi(t)=2 \sigma_0\JacobiAmplitude{k\omega_0 t+\Delta\phi}{k^{-2}}.
\label{eq:phi_exact_und}
\end{equation}
The term $\Delta\phi$ is a dephasing defined for any initial condition $\phi_0$ by:
\begin{equation}
\Delta\phi=\text{sn}^{-1}\left(\sin\left(\frac{\phi_0}{2}\right)\bigg| k^{-2}\right).
\label{eq:Delta_phi}
\end{equation}
Additionally, $\sigma_0$ determines the sign of $\phi(t)$ and is determined by $\dot{\phi}_0$ or equivalently by $n_0$. It is given by: 
\begin{equation}
\sigma_0=\left\{\begin{aligned}
         & \text{sign}(n_0)                         &\text{if } n_0\neq 0, \\
         & 1                                        &\text{if } n_0= 0.  \\
  \end{aligned}
  \right.
  \label{eq:sigma}
\end{equation}

The derivative of the Jacobi amplitude is the Jacobi-DN elliptic function \cite{abramowitz_handbook_1972}:
\begin{equation}
\frac{d\JacobiAmplitude{k\omega_0 t}{k^{-2}}}{dt}=k\omega_0\JacobiDN{k\omega_0 t}{k^{-2}}.
\label{eq:am_dot}
\end{equation}
Therefore, the imbalance evolution $n(t)$ is:
\begin{equation}
n(t)=\sigma_0 N_0\JacobiDN{k\omega_0 t+\Delta\phi}{k^{-2}},
\label{eq:n_exact_und}
\end{equation}
with $N_0$ the extremal value reached by the imbalance defined by:
\begin{equation}
N_0=\frac{2k}{\sqrt{\Lambda+\lambda}}.
\label{eq:N0}
\end{equation}
 
Eqs.~(\ref{eq:phi_exact_und},\ref{eq:n_exact_und}) describe all dynamical modes of the BJJ, which differ by the value of $k$. For $k=0$, the system is at equilibrium. For ${0<k<1}$, the dynamics present Josephson oscillations. For $k=1$, the dynamics follow the separatrix. Finally, ${1<k\leq (\sqrt{\Lambda+\lambda})/2}$ corresponds to the MQST. The upper boundary on $k$ comes from the limitation that the largest possible imbalance is $N_0 = 1$.

\subsection{Generalisation to an asymmetric BJJ}\label{sec:asymmetry}

\begin{figure}[h] 
\includegraphics[width=\columnwidth]{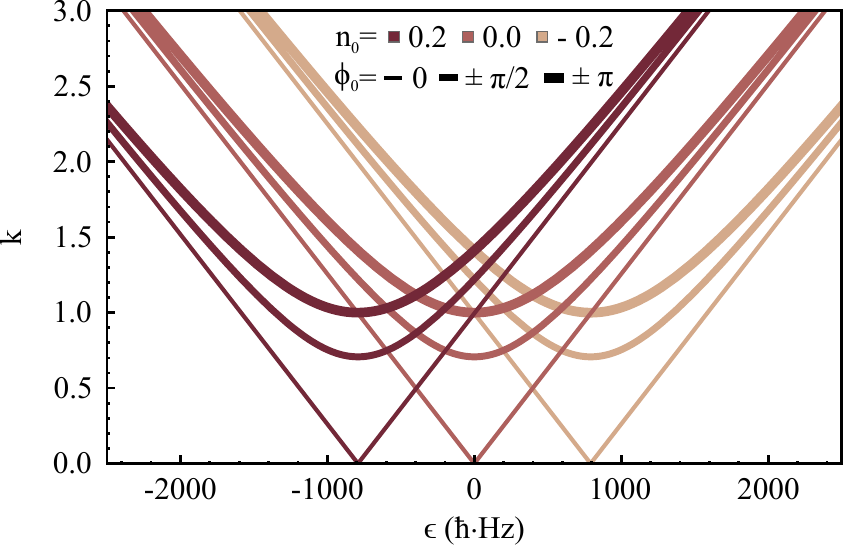}
\caption{\label{Fig:k_epsilon} Effect of the detuning $\epsilon$ on $k$ for 15 initial states. The trap is characterized by $J=\SI{20}{\hbar\cdot \mathrm{Hz}}$ and $\Lambda=100$. The detuning shifts the equilibrium position of the system in imbalance and changes the energy of the system.} 
\end{figure}

In the following, we consider an asymmetric trap such that $\epsilon\neq0$ (cf. Fig.~\ref{Fig:BJJ_schematics}). We assume that the asymmetry maintains the trap frequencies and that consequently it does not affect $J$ and $\Lambda$. The introduction of a detuning goes beyond the pendulum analogy. However, we can adapt the solutions defined previously to include a detuning of arbitrarily large values. For this we evaluate Eq.~(\ref{eq:phi_dot_sym}) at the initial time for $\epsilon\neq 0$:
\begin{align}
\dot{\phi}_0(\epsilon)&=\frac{\epsilon}{\hbar} + \frac{2J}{\hbar}\left(\Lambda+\lambda\right)n_0, \label{eq:phi_dot0_asym}
\end{align}
and interpret the detuning as an additional contribution to the kinetic energy. Consequently, the anharmonicity parameter $k$ now depends on $\epsilon$:   
\begin{equation}
k(\epsilon)=\frac{1}{2\omega_0}\sqrt{\dot{\phi}_0(\epsilon)^2 +4\omega_0^2\sin^2(\frac{\phi_0}{2})}. \label{eq:k_asym}
\end{equation} 
The effect of the detuning on the dynamics can be partly described by its effect on the anharmonicity. We illustrate this on Fig.~\ref{Fig:k_epsilon} for multiple initial conditions involving the different combinations of ${n_0=[-0.2,0,0.2]}$ and ${\phi_0=[-\pi,-\pi/2,0,\pi/2,\pi]}$. From Eq.~(\ref{eq:k_asym}), we deduce that $k$ reaches a minimum at ${\epsilon_L=-2J (\Lambda +\lambda)n_0}$. In the specific case of $\phi_0=0$, the system is at equilibrium as ${k(\epsilon_L)=0}$. It follows that the detuning shifts the equilibrium point in imbalance. 

We observe that the behaviour of $k$ presents several symmetries with respect to $\epsilon$, as illustrated by Fig.~(\ref{Fig:k_epsilon}). We first notice that $k(n_0,\pm\phi_0)$ and $k(-n_0,\pm\phi_0)$ are symmetric with respect to $\epsilon=0$. This is explained by the geometry of the double well. Indeed, the imbalance $n_0$ in an $\epsilon$-detuned trap is indistinguishable from the imbalance $-n_0$ in a $-\epsilon$-detuned trap. 

Additionally, $k$ presents a symmetry with respect to $\epsilon=\epsilon_L$  such that $k(\epsilon_L\pm\epsilon)$ has an identical value. This can be understood geometrically in the phase portrait representation: the initial state ($n_0,\phi_0$) is at equal distant to the two equilibrium points of the traps of detuning $\epsilon_L+\epsilon$ and $\epsilon_L-\epsilon$. However, the initial state in these two traps does not have the same dynamics as it evolves with respect to different equilibrium points. It shows that $k$ is not sufficient to describe the detuning.

We mentioned that $\epsilon$ shifts the imbalance value of the equilibrium position. Using Eq.~(\ref{eq:phi_dot_sym}), we define the imbalance offset  $\Delta n$ as:
\begin{equation}
\Delta n=-\frac{\epsilon}{2J(\Lambda+\lambda)}. \label{eq:delta_n}
\end{equation}
As a result, the imbalance obeys: 
\begin{equation}
n(t)=\sigma_0 N_0\JacobiDN{k(\epsilon)\omega_0 t+\Delta\phi}{k^{-2}}+\Delta n -\bar{n}, \label{eq:n_ex_asym}
\end{equation}
with $\sigma_0$ the sign function of $n_0-\Delta n$ and $\bar{n}$ the mean value of the imbalance defined by:
\begin{equation}
  \bar{n}=\left\{\begin{aligned}
        & 0                                       &k\leq1, \\
        & \frac{N_0(\sqrt{1-k^{-2}}+1)}{2}         &k>1. \\
  \end{aligned}
  \right.
  \label{eq:n_bar}
\end{equation}

In the MQST ($k>1$), the averaged value of $n(t)$ is deduced from the variations of the Jacobi DN function written in Eqs.~(\ref{eq:y_dot},\ref{eq:am_dot}). 
The term $\bar{n}$ is introduced to center the oscillations around zero. Indeed, the modification of $k$ by $\epsilon$ can lead to a transition between Josephson oscillations and MQST which introduces an offset $\bar{n}$ that we want to distinguish from $\Delta n$.

The phase evolution is described by Eq.~(\ref{eq:phi_exact_und}) given that we take onto account the new definitions of $k(\epsilon)$ and $\sigma_0(\Delta n)$.

\subsection{Connection to the TMBH model and interpretation}\label{sec:link_BJJ_pend_no_eta}
In this section, we express the parameters of the TMBH model as a function of the pendulum parameters. We can combine the definitions of $\omega_0$, $N_0$ and $k$ given by Eqs~(\ref{eq:omega_0},\ref{eq:N0},\ref{eq:k_asym}) to express $J$ as: 
\begin{equation}
J=\frac{\hbar \omega_0}{4|\sin(\phi_0/2)|}\sqrt{|N_0^2-4(n_0-\Delta n)^2|}. \label{eq:J_gene}
\end{equation}
Using the definition of $\omega_0$, it immediately follows that:
\begin{equation}
\Lambda=\frac{4\sin(\phi_0/2)^2}{|N_0^2-4(n_0-\Delta n)^2|}-\lambda, \label{eq:Lambda_gene}  
\end{equation}
from which we deduce $U$ using $\Lambda=NU/2J$. From Eq.~(\ref{eq:delta_n}) we establish:
\begin{equation}
\epsilon=-\frac{2\hbar|\sin(\phi_0/2)|\Delta n}{\sqrt{|N_0^2-4(n_0-\Delta n)^2|}}. \label{eq:epsilon_gene}
\end{equation}

For a small asymmetry, the dependence of $k$ on $\epsilon$ is negligible and the definitions simplifies to: 
\begin{align}
        J &=\frac{\hbar\omega_0}{2}\frac{N_0}{2k}, \label{eq:J} \\ 
        \Lambda &=\frac{4k^2}{N_0^2}-\lambda, \label{eq:Lambda}  \\
        \epsilon&=-\frac{2\hbar\omega_0 k}{N_0}\Delta n. \label{eq:epsilon}
\end{align}
 
The advantage of the pendulum model is the interpretation of its parameters and their experimental accessibility. 

$N_0$ is the highest imbalance reached by the system (in absolute value) and is directly seen on the imbalance dynamics, both in the oscillating and in the MQST regime. $\Delta n$ is a shift of the imbalance that can be easily obtained from the data.

For small $\epsilon$, the energy ratio $k$ obeys the equation:
\begin{equation}
k=\frac{N_0}{\sqrt{N_0^2-n_0^2}}\sin(\phi_0/2), \label{eq:k_init}
\end{equation}
if $n_0\neq N_0$. The visual interpretation of $k$ differs between the two regimes. In the oscillating regime, the definition of $k$ given by Eq.~(\ref{eq:k}) leads to $k=\sin(\Phi_0/2)$ where $\Phi_0$ is the amplitude of the oscillations in phase, easily accessible experimentally. In the MQST, $k$ appears in the amplitude of the imbalance oscillations. Following Eqs.~(\ref{eq:y_dot},\ref{eq:am_dot}), the imbalance varies (in absolute values) between $N_0$ and $N_0\sqrt{1-k^{-2}}$. The value of $k$ is then deduced directly from the amplitude of the imbalance oscillations. 

The frequency of the anharmonic oscillations with first order correction is:
\begin{equation}
\omega\approx\omega_0\left( 1-\frac{\Phi_0^2}{16}\right).
\end{equation}
As $\Phi_0$ is well determined by the oscillation amplitude, $\omega_0$ is very reliably derived from $\omega$. In the MQST, the plasma frequency $\omega_0$ appears is the slope of the phase accumulation as $2\omega_0 k$.

The pendulum-like solutions Eqs.~(\ref{eq:phi_exact_und},\ref{eq:n_exact_und}) allow a reliable estimation of $\Lambda$, $J$ and $\epsilon$. 

\subsection{Limitation of the analytical solutions}\label{sec:limitation}

The description of the BJJ with the analytical solutions of a rigid pendulum holds true as long as the maximal variation of the pendulum length ${l(t)}$ defined by ${l(t)=\sqrt{1-n^2(t)}}$ is negligible. 

In the oscillating regime, the variations of the imbalance are given by:
\begin{equation}
\Delta n \leq |n(t)| \leq N_0 + \Delta n.
\label{eq:var_n_JO}
\end{equation}
Consequently, the maximal variation of the pendulum length in the oscillating regime reads
\begin{equation}
\delta_\text{JO}=\sqrt{1-(N_0 + \Delta n)^2}-\sqrt{1-(\Delta n)^2}.
\end{equation} 

In the MQST regime, the amplitude of the imbalance oscillation is given by the variations of the Jacobi DN elliptic function. It follows from Eqs.~(\ref{eq:y_dot},\ref{eq:am_dot}) that:
\begin{equation}
N_0\sqrt{1-k^{-2}}+\Delta n\leq |n(t)| \leq N_0 +\Delta n.
\label{eq:var_n_MQST}
\end{equation}
Consequently, $\delta_\text{MQST}$ reads:
\begin{equation}
\delta_\text{MQST}=\sqrt{1-(N_0+\Delta n)^2}-\sqrt{1-(N_0\sqrt{1-k^{-2}}+\Delta n)^2}.
\end{equation} 

If $N_0$, $\Delta n$ and $(n_0,\phi_0)$ are known, the hypothesis of negligible momentum-shortening can be verified \textit{a posteriori}. 

\section{Dynamics of a dissipative Bosonic Josephson Junction}

\begin{figure}[!] 
\includegraphics[]{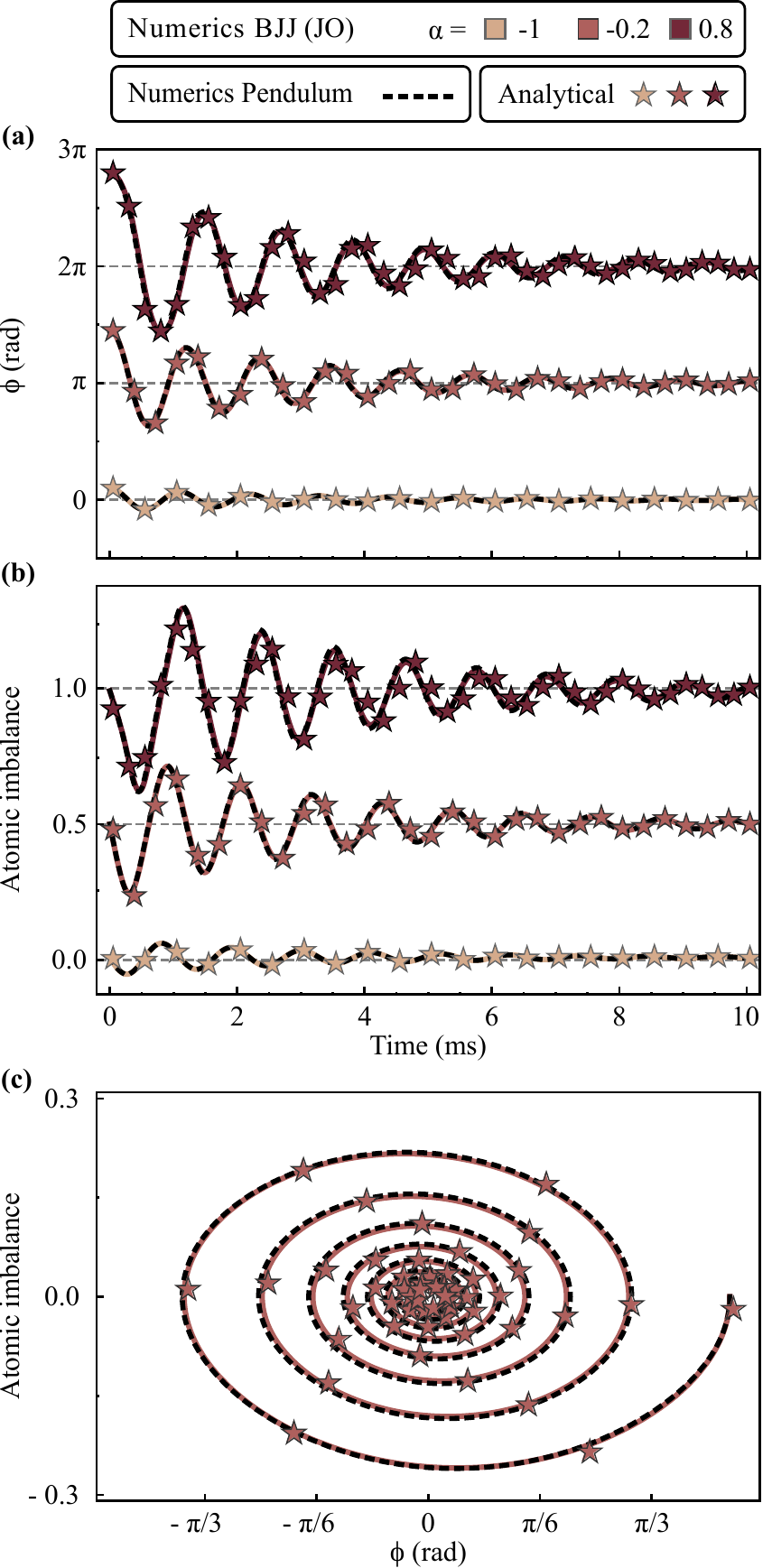}
\caption{\label{Fig:dampedJO}  Time evolution of the phase (a) and atomic imbalance (b) for $N=$\SI{5000}{atoms} in the trap characterized by $J=\SI{100}{\hbar\cdot \mathrm{Hz}}$, $U=\SI{0.8}{\hbar\cdot \mathrm{Hz}}$ and a viscosity $\eta=120$. The three initial states are chosen to initiate an oscillating dynamics ($\alpha<1$): $n_0=0$ and $\phi=[0.1,0.45,0.8]\Pi$ corresponding to $\alpha=-1,-0.2$ and $0.8$. For clarity, the phase evolution is shifted by a unit of $\pi$ and the imbalance is shifted by $0.5$ unit and their respective zero is indicated by a gray dashed line. Under the effect of $\eta$ the dynamics exponentially damp toward equilibrium. The various curves represent the numerical resolution of the equations of motion of the Bosonic Josephson Junction in the mean-field 2-site Bose-Hubbard model (color plain lines), the corresponding numerical resolution of the pendulum (black dashed line) and  the corresponding heuristic solutions (color stars). For clarity, the phase portrait (c) is showed for the second initial state only. The other initial states present a similar damping. } 
\end{figure}
\begin{figure}[!] 
\includegraphics[width=\columnwidth]{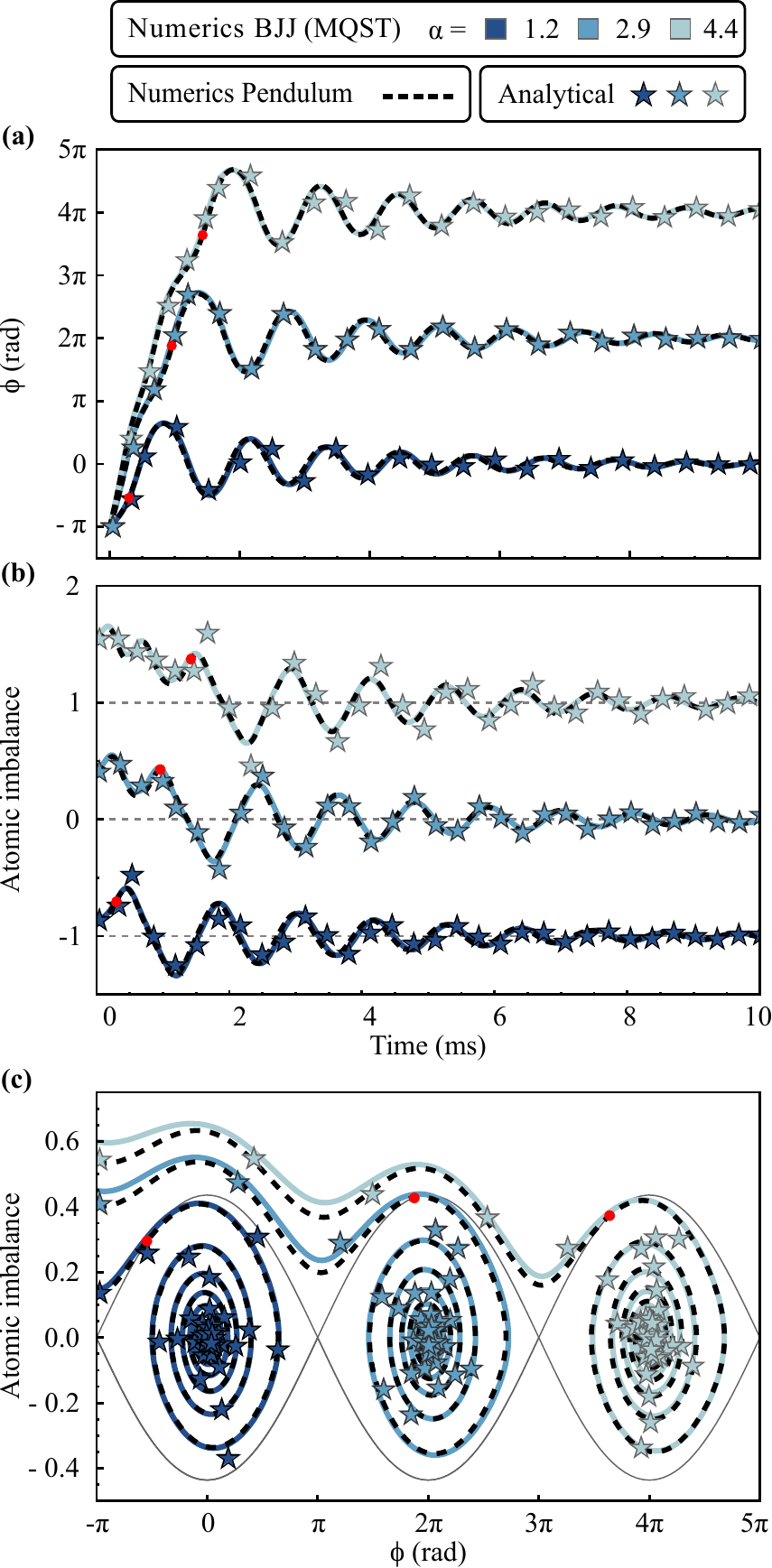}
\caption{\label{Fig:DampedMQST} Time evolution of the phase (a) and atomic imbalance (b) and phase portrait representation (c) for $N=$\SI{5000}{atoms} in the trap characterized by $J=\SI{100}{\hbar\cdot \mathrm{Hz}}$, $U=\SI{0.8}{\hbar\cdot \mathrm{Hz}}$ and a viscosity $\eta=120$.  The system is initialized in the self-trapping with the three initial states: ${n_0=[0.15,0.45,0.6]}$ and $\phi=-\pi$ corresponding to ${\alpha_{t=0}=1.4,4.1}$ and $8$. The imbalance evolution is shifted by one unit for each initial state for clarity and there respective zero is indicated by a gray dashed line. Under the effect of $\eta$, the system decays in the self-trapping regime until it reaches the separatrix at the time $\tilde{t}=\log(k)\tau$ (represented by a red dot). It then enters the oscillating regime and exponentially damps to reach equilibrium. We represent the numerical resolution of the equations of motion of the BJJ in the mean-field 2-site Bose-Hubbard model (color plain lines), the numerical resolution of the equivalent pendulum (black dashed line) and  the corresponding analytical solutions (color stars). }
\end{figure}

\subsection{Dissipation in the TMBH model}
The mean-field TMBH model is non-dissipative such that the dynamics remain undamped over time. However, \cite{marino_bose-condensate_1999} introduces a dissipative term proportional to $\dot{\phi}(t)$ in Eq.~(\ref{eq:n_dot_sym}) analogous to a pendulum with friction. This relies on a dimensionless viscosity $\eta$ normalized to the total atom number $N$ for consistency with the definition of the normalized imbalance $n$. The damped equations of motion of a BJJ read:
\begin{align}
\dot{n}(t)&=-\frac{2J}{\hbar}\sqrt{1-n^2(t)}\sin\phi(t)-\frac{\eta}{N}\dot{\phi}(t), \label{eq:n_dot_damp} \\
\dot{\phi}(t)&=\frac{\epsilon}{\hbar}+\frac{2J}{\hbar}\left[\Lambda n(t)+\frac{n(t)}{\sqrt{1-n^2(t)}}\cos\phi(t)\right]. \label{eq:phi_dot_damp}
\end{align}

To recover the pendulum analogy, we first consider a symmetric BJJ ($\epsilon=0$). Following a similar approach as for Eq.~(\ref{eq:phi_dot_dot}), we establish the pendulum-like equation in a dissipative medium: 
\begin{equation}
 \ddot{\phi}(t)+\frac{2}{\tau}\dot{\phi}(t)+\omega_0^2\sin(\phi(t))=0, \label{eq:phi_dot_dot_damp}
\end{equation}
where $\omega_0$ is the plasma frequency and $\tau$ is a characteristic decay time defined by:
\begin{equation}
\frac{2}{\tau}=\frac{2J}{\hbar} \frac{\eta}{N}(\Lambda+\lambda).
\label{eq:tau}
\end{equation}

The numerical agreement between the damped TMBH model and the damped pendulum is remarkable (see Fig.~\ref{Fig:dampedJO}, \ref{Fig:DampedMQST}), such that we extend the pendulum analogy to establish the heuristic solutions of a generic BJJ (i.e. for any value of $\epsilon$) with dissipation.

\subsection{Heuristic solutions of a generic BJJ with dissipation}
The analytical derivation presented in Sec.~[\ref{sec:analytical_sol_no_eta}] relies on energy conservation and cannot be done exactly with dissipation. As a result, the decay is heuristically determined by combining the results of the damped pendulum in the harmonic regime with the results of the ideal BJJ. 

In the harmonic regime, Eq.~(\ref{eq:phi_dot_dot_damp}) is exactly solvable. The oscillations have a sinusoidal shape of frequency
\begin{equation}
\omega(\tau)=\sqrt{\omega_0^2-\frac{1}{\tau^2}}, \label{eq:omega}
\end{equation}
and present an exponential damping characterized by $\tau$.

In the undamped analytical solutions, the energy appears in $k$,which makes it the relevant quantity to decrease over time. Its initial value is affected by $\tau$ through $\omega$: 
\begin{equation}
k_0(\epsilon,\tau)=\frac{1}{2\omega(\tau)}\sqrt{\dot{\phi}_0(\epsilon)^2 +4\omega(\tau)^2\sin^2(\frac{\phi_0}{2})}. \label{eq:k_asym_damp}
\end{equation} 

We then introduce an exponential damping of $k_0(\omega,\tau)$ characterized by $\tau$. Combining these results, we obtain the heuristic formula for the evolution of the phase of a dissipative BJJ: 
\begin{align}
\phi(t)=2\sigma_0\JacobiAmplitude{(\omega k_0 t+\Delta\phi) \text{e}^{-t/\tau}}{k_0^{-2}\text{e}^{2t/\tau}}, \label{eq:phi_damp} 
\end{align}
with $\Delta\phi$ the phase shift defined by:
\begin{equation}
\Delta\phi=\text{sn}^{-1}\left(\sin\left(\frac{\phi_0}{2}\right)|k_0^{-2}\right). \label{eq:delta_phi_damp}
\end{equation}

The exponential decay of $\Delta\phi$ is imposed by the convention of the Jacobi amplitude, which is equivalent to a sinusoidal function of frequency $\omega$ in the limit where the argument reaches zero. 

Eq.~(\ref{eq:phi_damp}) is in principle valid both in the self-trapping and in the oscillating regime. It describes accurately the oscillating regime up to $k\approx1$.
The decay decreases the amplitude of the oscillations exponentially and the frequency of the oscillation increases toward $\omega$ (see Fig.\ref{Fig:dampedJO}). 

However, in the vicinity of the separatrix, Eq.~(\ref{eq:phi_damp}) describes very well the exponential decay of the amplitudes, but does not evaluate correctly the change of frequency. As the numerical pendulum keeps matching very well the TMBH predictions, we deduce that the pendulum analogy remains true, but that the assumption of exponential decay derived from the harmonic regime must be adjusted. 

The oscillations exhibit a decay with two time-scales: a first one $\tau$ for the envelop and a second one $\tau_2$ for the frequency increase. To decouple these two time-scales, we decompose the Jacobi amplitude in the two functions it involves \cite{abramowitz_handbook_1972}. The envelop term is defined by a $2\arcsin(k_0\text{e}^{-t/\tau})$. The oscillating component results from the Jacobi SN function $\JacobiSN{\omega k_0 \text{e}^{-t/\tau_2} t}{(k_0 \text{e}^{-t/\tau_2})^{-2}}$. We normalize the Jacobi SN by its argument to maintain the oscillation between $-1$ and $1$. 

For large amplitude oscillations, we define the phase by: 
\begin{align}
\phi(t)=&\frac{1}{k_0 \text{e}^{-t/\tau_2}}\times 2\sigma_0\arcsin(k_0\text{e}^{-t/\tau}) \label{eq:phi_damp_JO}\\ 
        &\vspace*{0.2cm} \times\JacobiSN{(\omega k_0 t+\Delta\phi) \text{e}^{-t/\tau_2} }{k_0^{-2} \text{e}^{2t/\tau_2}}. \nonumber
\end{align}
Experimentally, the distinction between the two time-scales is negligible such that we treat $\tau_2$ as a correction to $\tau$. 

The decay in the MQST is very well described by Eq.~(\ref{eq:phi_damp}). It reproduces the exponential decrease of the phase accumulation until $k_0\text{e}^{-\tilde{t}/\tau}=1$ where the system reaches the separatrix. At this point, the Jacobi Amplitude presents a divergence. The time $t=\text{ln}(k_0)\tau$ must be excluded to avoid a non-physical divergence of the dynamics. Right after crossing the separatrix, the system undergoes large amplitude oscillations obeying Eq.~(\ref{eq:phi_damp_JO}) before we recover the behaviour predicted by Eq.~(\ref{eq:phi_damp}) (see Fig.\ref{Fig:DampedMQST}). 

The evolution of the imbalance has a general expression as function of $\dot{\phi}(t)$:
\begin{equation}
n(t)=\frac{N_0}{2\omega k_0}\dot{\phi}(t) +\Delta n,\label{eq:n_damp}
\end{equation}
with $\Delta n$ the imbalance at equilibrium defined by Eq.~(\ref{eq:delta_n}) and $N_0$ the amplitude of the undamped oscillations. 

\subsection{Connection to the TMBH model and interpretation}

Similarly to the ideal case, the parameters of the dissipative TMBH model can  be linked to the measurable parameters of the pendulum. 
The formulas for $J$, $\Lambda$ and $\epsilon$ remain true in the damped case. It follows that the viscosity $\eta/N$ can be expressed in term of measurable quantities using the definition of $\tau$ given by Eq.~(\ref{eq:tau}):
\begin{equation}
\frac{\eta}{N}=\frac{N_0}{k_0\tau\omega_0}.
\label{eq:eta}
\end{equation}
$\tau$ can be evaluated on data. Indeed, we expect the initial amplitude of the oscillations to decrease by 2 at the time $t=\tau \ln(2)$. The interpretation of $N_0$, $\omega_0$ and $k_0$ previously given in Sec.~[\ref{sec:link_BJJ_pend_no_eta}] must be adjusted to the damped case. 

$N_0$ is the undamped amplitude of the imbalance oscillations. It can be extracted from the damped imbalance oscillations at a time $\tilde{t}$ where they reach an extrema $N_0^d$. In this case the damped extrema is related to the undamped amplitude by $N_0^d=N_0 \text{e}^{-\tilde{t}/\tau}$. The frequency $\omega$  is obtained accurately after the system enters the harmonic regime and $\omega_0$ can be deduced using Eq.~(\ref{eq:omega}). $k_0$ is deduced in the oscillating regime from the maximum of the phase oscillation $\Phi_0^d$ reached at a time $\tilde{t}$ through $k=\sin(1/2\times\Phi_0^d\text{e}^{-\tilde{t}/\tau})$. In the MQST, it is linked to the slope at short time of the phase evolution which reads $k_0 \omega$. 

\section{Application to experimental data}\label{sec:data}

\begin{figure*}[!]
\includegraphics[]{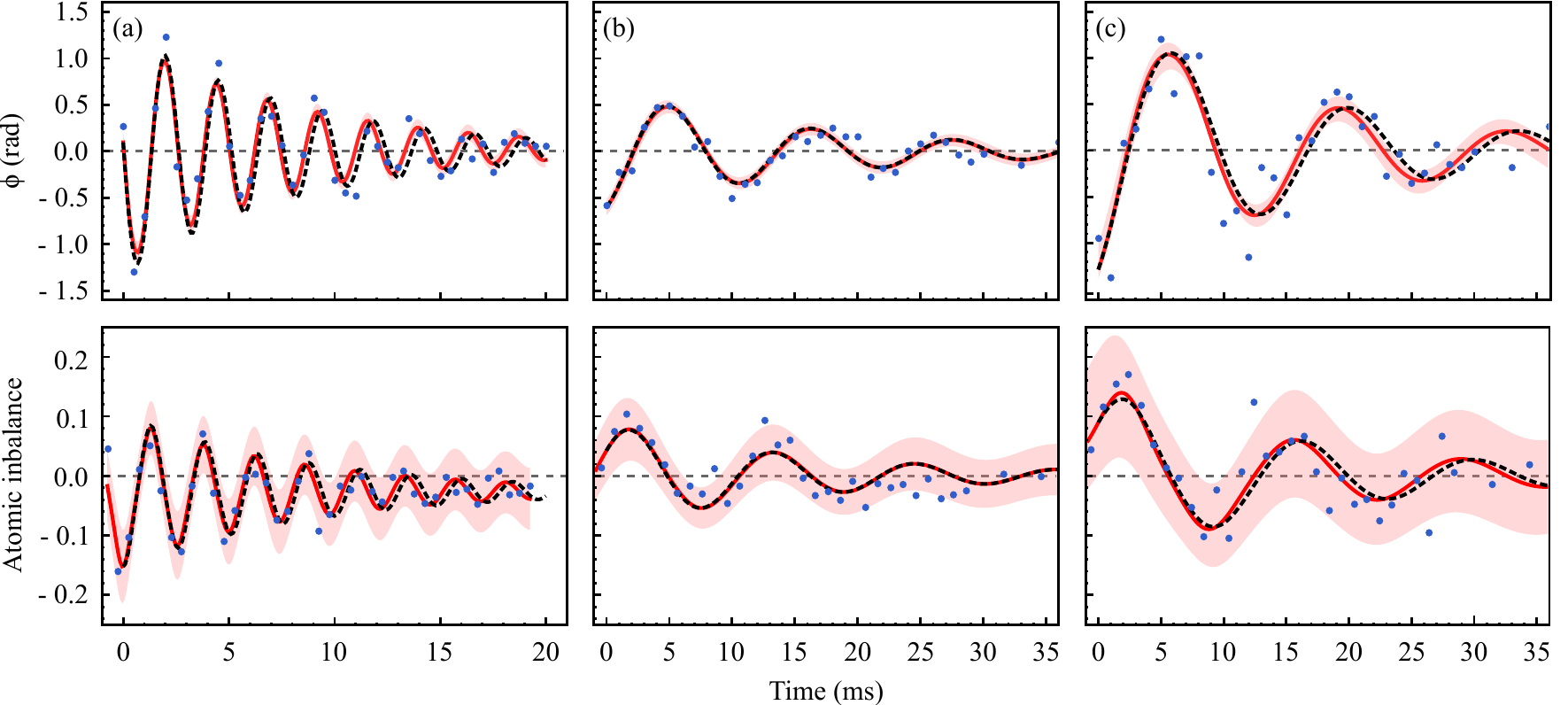}\\
\caption{\label{Fig:data_many_atoms} \textbf{Time evolution of the phase (top) and atomic imbalance (bottom) for various trap parameters and initial states.} The fit results are summarized in the Table \ref{Tab:parameters_values}. Blue dots: Experimental data. Red line: Fit result with heuristic functions. Black dashed line: Numerical resolution of the damped TMBH model with the parameters $J$, $\Lambda$ extracted from the heuristic fit. The plots show three different situations. (a) high atom number (3500 atoms) in an asymmetric trap ($\epsilon=\SI{103}{\hbar\cdot \mathrm{Hz}}$) with large tunnel coupling ($J=\SI{22}{\hbar\cdot \mathrm{Hz}}$); (b) small phase difference for a small atom number (750 atoms) and a symmetric trap with weak coupling ($J=\SI{6}{\hbar\cdot \mathrm{Hz}}$); (c) large initial phase for an atom number and trap geometry identical as for (b).} 
\label{Fig:fit_data}
\end{figure*}

In this last section, we apply the model to three sets of experimental data. The data are measured following the protocol detailed in \cite{pigneur_relaxation_2018}.

The system initially consists of a single 1D-BEC of $^{87}$Rb trapped in a magnetic trap generated with an atom chip \cite{folman_controlling_2000}. The typical trap parameters are ${\omega_{x,y}=2\pi\times\SI{3}{kHz}}$ and ${\omega_{z}=2\pi\times\SI{22}{Hz}}$. The trap contains typically few thousands atoms at the temperature ${T\approx\SI{20}{nK}}$. After condensation, the trap geometry is  modified by radio-frequency dressing to reach a symmetric and elongated double-well with a tunable barrier height \cite{schumm_matter-wave_2005}. The values of the trap frequencies are typically divided by two compared to the single trap. The splitting of the wave-packets is done adiabatically. At this stage, the phase difference is $\phi\approx 0$ with limited longitudinal excitations. The imbalance is $n\approx 0$ with a large number-squeezing factor of typically $\xi_N=0.57(6)$ \cite{berrada_integrated_2013, pigneur_relaxation_2018}. The relative phase is then imprinted in a trap of negligible coupling by introducing an energy difference between the two sites of the double-well. While the phase in a decoupled trap is expected to randomize under the effect of interaction-induced phase diffusion \cite{lewenstein_quantum_1996, javanainen_phase_1997}, this effect is reduced for a number-squeezed state. As a result, the prepared relative phase $\Phi_0$ presents small statistical fluctuations as illustrated by the phasor $R=0.91(2)$. After phase imprinting, the barrier is lowered to recouple the wave-functions. The tunneling dynamics starts during the recoupling stage, such that the initial phase and imbalance measured can differ from $\phi=\Phi_0$ and $n=0$. The imaging protocol is a time-of-flight measurement of either the phase or the imbalance. The detection is done by fluorescence with single-atom sensitivity \cite{bucker_single-particle-sensitive_2009}.   

In Fig.~\ref{Fig:fit_data}, we show the dynamics obtained with three different initial conditions. Fig.~\ref{Fig:fit_data}(a) shows the dynamics of a large atom number $N=\SI{3500}{\text{atoms}}$ oscillating in an asymmetric trap of low barrier height. Fig.~\ref{Fig:fit_data}(b),(c) show measurements obtained for significantly less atoms $N=\SI{750}{\text{atoms}}$ oscillating in a symmetric trap with a high barrier. They differ by the value of their initial phase difference. The initial phase in Fig.~\ref{Fig:fit_data}(b) is half of the one of Fig.~\ref{Fig:fit_data}(c).

The fit model is given by Eqs.~(\ref{eq:phi_damp},\ref{eq:n_damp}) and takes onto account a damping and an asymmetry. We consider $k, \omega_0, N_0, \tau, \Delta\phi$ and $\Delta n$ as fit parameters and report their values in Table~\ref{Tab:parameters_values}. 

\begin{table}[b]
\renewcommand{\arraystretch}{1.2}
\centering
       \begin{tabular}{cc c|c|c|c|c|c|c|c|c|c|c|c|}
            \cline{4-13}
			\multirow{2}{*}{}&\multirow{2}{*}{} &&  \multicolumn{6}{c|}{Fit parameters }&  \multicolumn{4}{c|}{Deduced TMBH}\\
	        \multirow{2}{*}{}&\multirow{2}{*}{} &&  \multicolumn{6}{c|}{derived from pendulum}&  \multicolumn{4}{c|}{parameters}\\	
			\cline{3-13} 
	     & &\multicolumn{1}{ |c|  }{$ N_\text{at}$} &  $k_0$& $N_0$& $\tau$ &$\omega_0$ &$\Delta n$&$\Delta \phi$& $\Lambda$ & $J/\hbar$ & $\epsilon/\hbar$ & $\eta$ \\			
	     & &  \multicolumn{1}{ |c|  }{}             &     &      &ms      & Hz      &          &            &           &  Hz       &Hz                 &       \\	\hline		
\hline 
   \multicolumn{1}{ |c  }{\multirow{2}{*}{(a)}}& \multicolumn{1}{ |c|  }{$\bar{x}$}   &3500 &0.57  &0.12  &8.9  & 2623 &-0.03 &-2.0     &92 & 22&103   &29   \\ \cline{2-13}
   \multicolumn{1}{ |c  }{}	                   & \multicolumn{1}{ |c|  }{$\sigma$}    &300  &0.03  &0.06  &0.9  &13    &0.02  & 0.14    &87  &  10  &92     &14    \\ \hline 
\hline
   \multicolumn{1}{ |c  }{\multirow{2}{*}{(b)}}& \multicolumn{1}{ |c|  }{$\bar{x}$}   &750  &0.31  &0.08  &17   &554   &0.01  &-0.6   &56 &6  &-0.8   &21   \\ \cline{2-13}
   \multicolumn{1}{ |c  }{}	                   & \multicolumn{1}{ |c|  }{$\sigma$}    &150  &0.03  &0.05  &3    &10    &0.01  & 0.09  &65  &3  &8     &12    \\ \hline 	
\hline
   \multicolumn{1}{ |c  }{\multirow{2}{*}{(c)}}& \multicolumn{1}{ |c|  }{$\bar{x}$}   &750  &0.60  &0.15  &17   &465   &0.001 & -0.7   &64 &5  & -0.5  &19   \\ \cline{2-13}
   \multicolumn{1}{ |c  }{}	                   & \multicolumn{1}{ |c|  }{$\sigma$}    &150  &0.06  &0.09  &3    &9     &0.03  &  0.08  &79 &3  & 22    &12    \\ \hline 
	\end{tabular}

\caption{Summary of the mean values $\bar{x}$ and standard error $\sigma$ for the measured atom number, for the fit parameters obtained with the heuristic model and deduced parameters of the damped mean-field TMBH model associated to the data sets of Fig.~(\ref{Fig:fit_data}).\label{Tab:parameters_values}}
\end{table}

To investigate the correlations between parameters obtained for the first set of data, we compute the correlation matrix $\mathscr{C}$ from the covariance matrix $C$ as detailed in Appendix~A. The correlation matrix obtained from the heuristic model presented in this study reads
  
  \[
\mathscr{C}_\text{heur} \approx \hspace{0.2cm}
\begin{blockarray}{c c c c c c }
\hspace{0.3cm}k & \hspace{0.3cm}\omega_0  & \hspace{0.3cm}\Delta\phi & \hspace{0.3cm}N_0 & \hspace{0.2cm}\tau&\hspace{0.2cm}\Delta n \\
\begin{block}{(r r r r r r )}
   \vspace{-0.2cm}\\
  1.00   & -0.09 &   0.02  &   0.07  &  -0.65  &  -0.01  \\
 -0.09   &  1.00 &  -0.47  &  -0.06  &   0.52  &   0.01  \\
  0.02   & -0.48 &   1.00  &   0.00  &   0.03  &  -0.01  \\
  0.07   & -0.06 &   0.00  &   1.00  &  -0.11  &  -0.02   \\
 -0.65   &  0.51 &   0.03  &  -0.11  &   1.00  &   0.01   \\
 -0.01   &  0.01 &  -0.01  &  -0.02  &   0.01  &   1.00 \\
  \vspace{-0.2cm}\\
\end{block}
\end{blockarray}
 \]

with the off-diagonal terms giving the correlations between parameters. 

To fit the data with the equations of motion of the damped TMBH model \cite{marino_bose-condensate_1999}, we consider $J, \eta, \Lambda, \epsilon, n_0$ and $\phi_0$ as fit parameters. The correlation matrix obtained using Eqs.~(\ref{eq:n_dot_damp},\ref{eq:phi_dot_damp}) as fit functions reads: 

  \[
\mathscr{C}_\text{TMBH} \approx \hspace{0.2cm}
\begin{blockarray}{c c c c c c }
\hspace{0.3cm}J  & \hspace{0.3cm}\eta & \hspace{0.3cm}\Lambda & \hspace{0.3cm}\epsilon & \hspace{0.2cm}n_0 & \phi_0 \\
\begin{block}{(r r r r r r )}
   \vspace{-0.2cm}\\
 1.00  &  0.82   &  -1.00  & -0.78  &  -0.34  &  0.00   \\
 0.82  &  1.00   &  -0.82  & -0.61  &  -0.06  & -0.42  \\
  -1.00  & -0.82   &   1.00  &  0.78  &   0.32  & -0.01  \\
   -0.78  & -0.61   &   0.78  &  1.00  &   0.05  &  0.01   \\
 -0.34  & -0.06   &   0.32  &  0.05  &   1.00  & -0.66  \\
  0.00  & -0.42   &  -0.01  &  0.01  &  -0.66  &  1.00 \\
  \vspace{-0.2cm}\\
\end{block}
\end{blockarray}
 \]

We notice that the correlations between $J,\eta, \Lambda$ and $\epsilon$ are significantly larger than in the heuristic model. This is similarly observed on the other data sets. 

We can now establish the connection between the heuristic model and the parameters of the damped TMBH model. Using the fit values of Table~\ref{Tab:parameters_values}(a) and  Eqs.~(\ref{eq:delta_phi_damp},\ref{eq:k}), we deduce the starting conditions ${\phi_0=\SI{0.07(4)}{rad}}$ and $n_0=0.12(4)$. Due to the tunneling dynamics during the recoupling stage, these values differ from the ones of the prepared state (maxima of the phase difference and zero imbalance). 

The parameters of the damped TMBH model are deduced from the fit values of Table~\ref{Tab:parameters_values}(a) and Eqs.~(\ref{eq:J},\ref{eq:Lambda},\ref{eq:epsilon},\ref{eq:eta}). We obtain ${J=\SI{22(5)}{\hbar\cdot \mathrm{Hz}}}$, ${\Lambda=92(43)}$, ${\epsilon=\SI{103(46)}{\hbar\cdot \mathrm{Hz}}}$, ${\eta=32(7)}$. An estimation of the error is done as detailed in  Appendix~A and compared to the damped TMBH model. 
The fit using the damped TMBH model leads to compatible results: ${\phi_0=\SI{0.07(20)}{rad}}$, ${n_0= 0.10(1)}$, ${J=\SI{19(3)}{\hbar\cdot \mathrm{Hz}}}$, ${\Lambda=128(34)}$, ${\epsilon=\SI{142(26)}{\hbar\cdot \mathrm{Hz}}}$, ${\eta=26(5)}$. However, the large correlations between parameters makes the fit very sensitive to the initial guess. 

Finally, we compare the results obtained for the three different data sets. We see that the second example has the smaller value of the rescaled initial energy $k$. This is explained by the low barrier height, small atom number and small initial phase, resulting in a small initial energy in the system. The third example has the largest $k$ due to its high initial phase. We expect the fit results of (b) and (c) to give identical values of the plasma frequency and imbalance offset $\Delta n$ as the trap geometry is unchanged. Using our model, this is indeed the case, resulting in similar $\Lambda, J$ and $\epsilon$ without any constrain on the system. The fit results using the damped TMBH model show very different results between the two data sets due to the convergence of the fit to local minima. 

\section{Conclusion}
In conclusion, we have shown that the analytical solutions of a rigid pendulum expressed in term of Jacobi elliptic functions describe the oscillating and self-trapped dynamics of a Bosonic Josephson Junction very accurately. We have established the range of validity of the model by defining a criteria linking the initial state of the system and the amplitude of the imbalance oscillations. Going beyond the pendulum analogy, we took into account the effect of an asymmetry in the double-well.

The predictions of the model are in very good agreement with the damped TMBH model presented by \cite{marino_bose-condensate_1999}. By drawing the connection between the two models, we gained insights about the physics of the TMBH model. Indeed, the pendulum model involves experimentally accessible parameters, such as the frequency and amplitude of the oscillations and we expressed the parameters of the TMBH model as a function of uncorrelated quantities of the system.

Motivated by the need of our experimental observation, we have established an heuristic formula describing a dissipation in a Bosonic Josephson Junction. We applied our model to experimental data and showed that the heuristic model gives reliable estimates of the parameters of the system, and in particular of the damping that needs to be understood.    

\section*{Acknowledgements}
We are grateful to I. Mazets, I. Lovas, G. Zaránd and E. Demler for helpful discussions.  M.P. acknowledges the support of the Doctoral Program CoQuS. This research was supported by the ERC advanced grant QuantumRelax and by the Austrian Science Fund (FWF) through the project SFB FoQuS (SFB F40).

\section*{Appendix A: Correlation matrix and error estimations}\label{App:error}
The solver used to fit the data does not directly provide the covariance matrix. Instead, it returns both the goodness-of-fit parameter $R$ and the Jacobian matrix $\mathcal{J}$ for the optimized fit parameters.

The goodness-of-fit parameter $R$ is the defined by
\begin{equation}
R=\sum_{y_\text{data}} (y_\text{model}-y_\text{data})^2,
\end{equation}
with $y_\text{model}$ the evaluation of the fit function at $x_{data}$ for the best fit parameters. 
Calling  $N_{data}$ the number of data and $\nu$ the number of fit parameters we compute the mean-squared error
\begin{equation}
\text{MSE} = \frac{R}{N_{data}-\nu}.
\end{equation}
The covariance is calculated by:
\begin{equation}
C=(\mathcal{J}'\times \mathcal{J})^{-1}\times\text{MSE}.
\end{equation}
The correlation matrix $\mathscr{C}$ is obtained by:
\begin{equation}
\mathscr{C}= D^{-1}\times C \times D,
\end{equation}
with $D$ containing the square-root of the diagonal elements of $C$.\\

The error on the TMBH model parameters $\Lambda,J,\epsilon$ and $\eta$ is estimated by propagation of errors derived by Taylor series expansion limited to the gradient term. For a multi-variable function with correlated parameters, the error estimation must also account for the cross-terms involving products of uncertainties between the various combinations of two variables. 
The general formula of the error $\sigma_f$ for a multi-variable $f(x_1,x_2...,x_N)$ reads:
\begin{equation}
\sigma^2_f=\sum_{i=1}^N \sigma_i^2 \left( \frac{\partial f}{\partial x_i}  \right)^2+2\sum_{i=1}^N\sum_{j\neq i}^N  \left( \frac{\partial f}{\partial x_i}\right)\left( \frac{\partial f}{\partial x_j}\right) \sigma_{ij},
\end{equation}
with $\sigma_i$ the variance of $x_i$ (i.e the coefficient $C(i,i)$ of the covariance matrix) and $\sigma_{ij}$ the covariance between $x_i$ and $x_j$ (i.e the coefficient $C(i,j)$ of the covariance matrix)

We illustrate it this formula $J(\omega_0,N_0,k)$ defined by Eq.~(\ref{eq:J}):
\begin{equation}
\sigma_J=\frac{\hbar\omega_0 N_0}{k} 
         \sqrt{
         \sum_{i=\omega_0,N_0,k}\frac{\sigma_i^2}{i^2}
            + 2\left(\frac{\sigma_{\omega_0 N_0}}{\omega_0 N_0}
            - \frac{\sigma_{k\omega_0}}{k\omega_0}
            - \frac{\sigma_{kN_0}}{kN_0} \right)
            }.
\end{equation}

In the example of the strong coupling data detailed in the paper and displayed in Fig.~\ref{Fig:fit_data}(a), the relevant coefficients of the covariance matrix are ${\sigma_{\omega_0}=13}$, ${\sigma_{N_0}=0.06}$, ${ \sigma_k=0.03}$, ${\sigma_{\omega_0 N_0}=-0.06}$, ${\sigma_{k\omega_0}=-0.1}$, ${\sigma_{kN_0}=0.0003}$, we obtain ${\sigma_J=\SI{11}{\hbar\cdot \mathrm{Hz}}}$. This error is entirely dominated by the error on $N_0$. This is also the case for $\Lambda$. For $\epsilon$, whose definition involves both $N_0$ and $\Delta n$ the error primarily comes from these two contributions. The small amplitude of the imbalance is the main source of error on the fit of the data.


\bibliography{model_library}

\clearpage
\onecolumngrid

\renewcommand{\thefigure}{S\arabic{figure}}
\setcounter{figure}{0}






\end{document}